\documentclass[12pt]{article}
\usepackage{epsfig}
\newcommand{\be}{\begin{equation}}
\newcommand{\ee}{\end{equation}}

\newcommand{\cc}{\cite}
\newcommand{\ba}{\begin{eqnarray}}
\newcommand{\ea}{\end{eqnarray}}
\newcommand{\bg}{\begin{gather}}
\newcommand{\foma}{\end{gather}}

\newcommand{\e}{\mbox{e}}

\newcommand{\vecc}[1]{\mbox{\boldmath $#1$}}

\newcommand{\sivfa}[1]{f_{1\,T}^{\perp\, #1}(x,k_{\perp})}

\def\e{\epsilon} 

\def\lagr{\hbox{$\cal L$}}

\def\f{\phi}
\def\F{\Phi}

\def\pb{\bar \psi}

\def\ex{\hbox{e}}

\def\F{\Phi}
\def\<{\langle}
\def\>{\rangle}

\def\a{\alpha}
\def\b{\beta}
\def\g{\gamma}  \def\G{\Gamma}
\def\d{\delta}  
   
\def\s{\sigma}
\def\r{\rho}  
\def\x{\xi}
\def\c{\chi}
\def\m{\mu}
\def\n{\nu}

\def\k{\kappa}

\def\({\left(}
\def\[{\left[}
\def\){\right)}
\def\]{\right]}

\def\cos{\hbox{cos}}
\def\sin{\hbox{sin}}
\def\inf{\infty}

\def\Tr{\hbox{Tr}}
\def\pa{{\cal P}}

\def\vkp{{\vecc k}_\perp}
\def\intz{\int\!\frac{d^3\vecc{k}}{(2\pi)^3}}
\def\into{\int\!\frac{d^3\vecc{k}_1}{(2\pi)^3}}

\def\intth{\int\!\frac{d^3\vecc{k}_3}{(2\pi)^3}}

\def\intksi{\int\!\frac{d\x^-d^2\x_\perp}{(2\pi)^3}}

\newcommand{\ct}[1]{\cite{#1}}

\begin{document}
\begin{titlepage}
%\hfill{SNUTP-05-011}\\

\begin{center}
{\Large \bf Instanton Contribution to the Sivers Function}

\vskip 5mm
I.O. Cherednikov$^{a, b, c ,}$\footnote{{\tt igor.cherednikov@jinr.ru}; {\it Alexander von Humboldt Fellow} } , U. D'Alesio$^{b
,}$\footnote{{\tt umberto.dalesio@ca.infn.it}}, N.I. Kochelev$^{c ,d ,}$\footnote{{\tt
kochelev@theor.jinr.ru}} , F. Murgia$^{b , }$\footnote{\tt francesco.murgia@ca.infn.it}

\vskip 5mm

$^a$ {\it Institut f\"{u}r Theoretische Physik II, Ruhr-Universit\"{a}t Bochum \\
D-44780 Bochum, Germany}\\
$^b$ {\it Dipartimento di Fisica, Universit\`a di Cagliari and INFN, Sezione di Cagliari \\
C.P. 170, I-09042 Monserrato (CA),
Italy}\\
$^c$ {\it Joint Institute for Nuclear Research \\ Bogoliubov
Laboratory of Theoretical Physics, Moscow Region, 141980
Dubna, Russia}\\
$^d$ {\it School of Physics and Astronomy and Center for
Theoretical Physics,
Seoul National University \\ Seoul 151-747, Korea }\\
\end{center}

 \vskip 0.5cm \centerline{\bf Abstract}

We study the Sivers function for valence $u$ and $d$ quarks in the proton within the
instanton model for QCD vacuum, adopting the MIT bag model wave functions for quarks. Within
approaches based on perturbative one-gluon final state interactions a non-zero value of the
Sivers function is related to the presence of both $S$ and $P$ wave components in quark wave
functions. We show that the instanton-induced chromomagnetic, nonperturbative interaction
leads to very specific spin-spin correlations between the struck and spectator quarks,
resulting in a non-trivial flavour dependence of the Sivers function. Comparison of the
obtained Sivers functions with phenomenological parameterizations is discussed.
\end{titlepage}
\setcounter{footnote}{0}

 \vskip 8mm

\section{Introduction}

Understanding the spin structure of the nucleon is a long-standing, challenging problem for
strong interaction theories which has received renewed interest in recent years. The
investigation of mechanisms able to explain the unambiguous observation of large transverse
single spin asymmetries (SSA) in inclusive meson production in hadronic collisions and in
semi-inclusive deep inelastic scattering (SIDIS) may help in understanding the origin of spin
effects in hadronic physics. Recently, significant SSA's in pion electroproduction on both
longitudinally \cite{HERMES1}, and transversely \cite{HERMES2} polarized proton targets have
been observed by the HERMES Collaboration at DESY. Analogous measurements and processes are
actively under investigation by the COMPASS experiment at CERN \cite{compass}. Measurements
of the SSA for pion production in polarized proton-proton collisions, both in the central
rapidity region and at large Feynman $x_F$, moderately large transverse momentum and center
of mass energy $\sqrt{s}=200$ GeV, have been performed by the PHENIX \cite{phenix}, STAR
\cite{star} and BRAHMS \cite{brahms} Collaborations at the Relativistic Heavy Ion Collider
(RHIC) at Brookhaven. These experiments plan to refine their measurements, improving the
statistics and enlarging the kinematical region explored.

We should emphasize that in spite of the fact that large, ``anomalous", SSA at high energy
and large transverse momentum were discovered long time ago in both inclusive and exclusive
hadron production in hadron-hadron collisions \ct{E704}, their understanding within QCD
\cite{siv90,coll93,qiu91}, although significantly improved in recent years
\cite{brodsky,coll02}, is far from being complete, see e.g. Ref.~\cite{efremov2}.

In perturbative QCD (pQCD) approaches with inclusion of intrinsic (or primordial) parton
motion, SSA can be related to the so-called naively T-odd, transverse momentum dependent
(TMD), Sivers distribution and Collins fragmentation functions \ct{mulders,ans95,sidis}.
These functions provide very important information on the behaviour of strong interactions at
large distances and can be calculated only within nonperturbative QCD. Therefore, in the
absence of first-principles QCD calculations for these functions, e.g., within lattice QCD,
it is very important to calculate SSA within some suitable model accounting for
nonperturbative QCD effects. In the last years it has been shown that the instanton model
(IM) can well describe the main nonperturbative aspects of QCD
\cite{shuryak,diakonov,schrempp,dorokhov,dch}. The first attempts to apply the IM to the
study of SSA were done in Ref.s~\cite{kochelev,kochelev1,shuryak1,hoyer}. In particular,  the
instanton induced gluon-quark interaction was shown to be important for the appearance of
significant SSA in quark-quark scattering \cite{kochelev,kochelev1}. In Ref.~\cite{hoyer}
this mechanism has been applied to estimate the current jet single spin asymmetry in SIDIS,
within a modified version of the final-state interaction approach suggested in
Ref.~\cite{brodsky}. The contribution to the SSA in SIDIS coming from the instanton-induced
spin-flip photon-quark vertex is the subject of a recent paper \cite{shuryak1}.

In this paper we derive  for the first time the Sivers distributions for $u$ and $d$ valence
quarks inside a proton, within  the instanton model for  QCD vacuum and using the MIT bag
model for quark wave functions. Part of the calculation is performed along the lines of
Ref.~\cite{YUAN}, where however the perturbative one-gluon exchange contribution to the
Sivers function was considered. Analogous calculations, using a quark-diquark model for the
nucleon, were presented in Ref.s~\cite{gamberg,bacchetta}.

\section{Sivers function within the MIT bag model formalism}

The Sivers function \cite{siv90}, $f_{1T}^{\perp\,\alpha}(x,\vkp^2)$ or
$\Delta^{\!N}f_{\alpha/p^\uparrow}(x,\vkp^2)$ (see Ref.~\cite{trento} for details on
different notations adopted in the literature), describes the correlation between the
intrinsic transverse momentum of an unpolarized quark (with flavour $\alpha)$  and the parent
proton transverse polarization. The general correlator for transverse momentum dependent
quark distribution functions is given by \ct{mulders} \be \F^{\alpha} (x, \vkp, S) = \int \!
\frac{d\x^-d^2\x_\perp}{(2\pi)^3} \ e^{-ik\cdot\x} \ \< P,S | \ \pb^{\alpha} (\x)
V^\dagger(\x) V (0) \psi^{\alpha}(0) \ | P,S \> |_{\x^+ =0} \ , \label{cf1} \ee where
$k^{\mu} = x P^{\mu}+k_\perp^{\mu}$ is the quark four-momentum, and $P, S$ are the hadron
momentum and spin vector respectively. The Wilson lines, also called gauge-links, $V(\xi)$,
are path-ordered exponentials required to preserve gauge invariance; as it was shown
recently, they are also crucial in order to get non-vanishing $T$-odd TMD parton
distributions, like the Sivers function \cite{brodsky,coll02}. They are defined as follows:
 \be V(\x) = \pa
\exp\(-ig_s \int_{\x^-}^\inf \! dz^-\,\hat{A}^+ (z^-, \x_\perp) \) \cdot \pa \exp\(-ig_s
\int_{\x_\perp}^\inf \! dz_\perp\,\hat{A}_{\perp} (z^- =\inf, z_\perp) \) \ , \label{linkx}
\ee
 where $g_s$ is the strong coupling constant, $\hat{A^\mu}=\sum_a t^a\,A_a^\mu$,
$t^a=\lambda^a/2$ are the Gell-Mann colour matrices and $A_a^{\mu}$ are the gluon gauge
fields. In the following, we will work in a covariant gauge, so that we can neglect the
second term in Eq.~(\ref{linkx}), which vanishes in nonsingular gauges \cite{ji_yuan}.

TMD distribution functions can be obtained by first opportunely Dirac-projecting and then
taking the trace of the general correlator, Eq.~(\ref{cf1}). In the case of interest, for
example, the Sivers function for a quark with flavour $\alpha$ can be extracted as follows:

\be \sivfa{\alpha} = \frac{1}{2}\,\frac{M_P}{2\,k_y}\ \Tr
\[\, \g^+ \,[\F^\a(x, \vkp, S_{\perp\,x}) - \F^\a(x, \vkp, -S_{\perp\,x})] \,\] \ , \label{sivf1}
\ee
 where $k_{\perp}=|\,\vecc{k}_{\perp}|$, $M_P$ is the proton mass and,
to be definite, we have considered a proton moving along the $+z$ direction and transversely
polarized along the $x$ direction; in this way, we have also subtracted the $\vkp$-even
contribution. Due to translational invariance, using Eq.~(\ref{cf1}) one can rewrite the last
equation in the form \be \sivfa{\alpha} = \frac{M_P E_P}{k_y} \ \intksi\ \ex^{-i k\cdot \xi }
\int\! d^3 \vecc r \ \<\,\uparrow_x|\, \bar \psi^\a (\x+ \vecc r) V^\dag (\x+\vecc r) \g^+ V
(\vecc r) \psi^\a (\vecc r)|\uparrow_x\,\> \ . \label{basic} \ee

Hereafter, for shortness, we will indicate by
 $\<\,\uparrow_x|\cdots|\uparrow_x\,\>$ any matrix element of the form
$[\,\<S_{\perp\,x}|\cdots|S_{\perp\,x}\>-\<-S_{\perp\,x}|\cdots|-S_{\perp\,x}\>\,]/2$.

In order to estimate the instanton contribution to the Sivers function, following
Ref.~\cite{YUAN} we will use (valence) quark wave functions from the MIT bag model. The
Fourier transform of the lowest modes of these functions, to which we limit, is then defined
as \be \psi^{\a,i}(\x) = \sum_m \intz \ \f_m({\vecc k}) \ \ex^{-ik \cdot \x} \ a^{i}_{\a,\,m}
\  , \label{psi} \ee
 where $a^{i}_{\a,\,m}$ is the quark annihilation operator, $\a$, $i$, are respectively the quark
flavour and colour indexes and $m$ is its spin component (helicity), $m=\pm 1/2$. Since we
consider only valence quarks in the proton, a similar term related to antiquark operators is
omitted. Moreover, in the MIT bag model
  \be \f_m ({\vecc k}) = N \sqrt{4\pi}\,
R_0^3 \ \({t_0 (k)\ \c_m}\atop{\vecc \s\!\cdot\!\hat{\vecc k}\ t_1(k)\ \c_m} \) \ , \ee
\noindent where $k=|\,\vecc{k}\,|$, $\hat{\vecc{k}}=\vecc{k}/k$,
 \be t_i (k) = \int_0^1 \! dx \ x^2 j_i(x k R_0) j_i (x\omega) \ , \ N^2 =
\frac{1}{R_0^3}\ \frac{\omega^3}{2 (\omega-1)\ \sin^2\ \omega} \  , \ee
 $R_0$ is the bag (confinement) radius, and $\omega=2.043$ for the lowest state of massless quarks
inside the bag. In the proton rest frame $E_P=M_P$ and the bag radius $R_0=4\omega/E_P$ is
fixed by the bag stability condition, $dE_P/dR=0$.

The transversely polarized proton ( e.g., in the $\pm\,x$-direction) is described by the
state vector \be |S_{\pm x}\> = \frac{1}{\sqrt{2}} \ \(|S_{+z}\> \pm |S_{-z}\>\) \ ,
\label{xpol1} \ee where the longitudinally polarized states read
 \begin{eqnarray} |S_{+z}\> \!\!&=& \!\!\frac{1}{\sqrt{6}}\,\frac{1}{\sqrt{18}} \,
\e_{ijk}\, \e_{n_1,n_2,n_3} \[\,a^{i\,\dag}_{u^\uparrow}(n_1)a^{j\,\dag}_{u^\uparrow}(n_2)
a^{k\,\dag}_{d^\downarrow}(n_3) -
a^{i\,\dag}_{u^\uparrow}(n_1)a^{j\,\dag}_{u^\downarrow}(n_2)a^{k\,\dag}_{d^\uparrow}(n_3)
\]\!|\,0\> \ , \\
 |S_{-z}\> \!\!&=& \!\!\frac{1}{\sqrt{6}}\frac{1}{\sqrt{18}} \, \e_{ijk}\, \e_{n_1,n_2,n_3}\
\[\,-a^{i\,\dag}_{u^\downarrow}(n_1)a^{j\,\dag}_{u^\downarrow}(n_2)
a^{k\,\dag}_{d^\uparrow}(n_3) +
a^{i\,\dag}_{u^\uparrow}(n_1)a^{j\,\dag}_{u^\downarrow}(n_2)a^{k\,\dag}_{d^\downarrow}(n_3)
\]\!|\,0\> \ ,\nonumber \end{eqnarray}
and the $n_i$ are position indexes.

\section{Instanton contribution}

The contribution to the quark Sivers function coming from perturbative, one-gluon exchange,
final state interactions (see Fig.~1a) has been considered in Ref.~\cc{YUAN} in the framework
of the MIT bag model. The interaction Lagrangian in this case is simply given by
 \be {\cal L}_{int}^{P} = - \sum_{\beta}\sum_{a,k,l}\, g_s\,
 \bar{\psi}^{\beta,k}(\eta)\,\gamma^{\mu}\,t^{a}_{kl}\,\psi^{\beta,l}(\eta)\,A_{a\,\mu}(\eta)\,
 \label{lintP}
 \ee
 where $\beta$ and $a$, $k$, $l$ are respectively flavour and colour indexes.
 It has been shown in Ref.~\cite{YUAN} that the interference between $S$ and $P$ wave
components of the bag quark wave functions leads to a non-zero value of the Sivers function.
Clearly, these contributions are related to the dynamics of quark confinement at the scale of
a typical hadronic size; as such, they can be responsible for the SSA at relatively small
transverse momentum, $k_\perp\approx 1/R_0\approx 200-300$ MeV/$c$.

In this paper we consider the contribution to the Sivers function which stems from a
completely different QCD dynamics (see Fig.~1b), namely from the nonperturbative
chromomagnetic quark-gluon interaction induced by  instantons \ct{kochelev}
\begin{figure}[h]
%\centering
%\vspace*{1.2cm}
%\centering
\hspace*{0.5cm} \epsfig{file=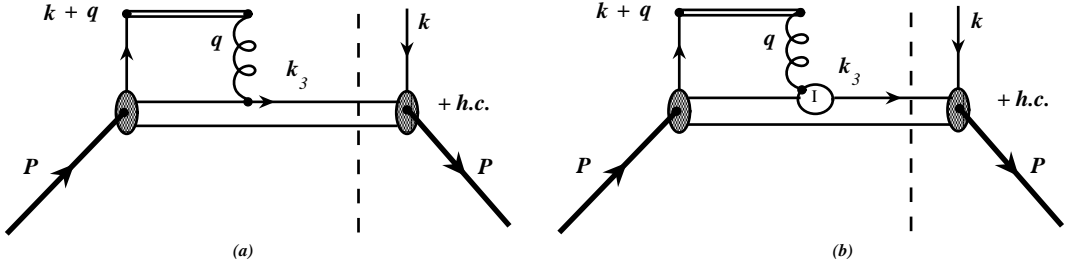,width=15cm,height=5cm}
%\vskip -8cm
\caption{Perturbative (a)  and instanton induced (b) contributions to the  Sivers function.
The  symbol $I$  denotes the instanton.}
%\vspace*{-5cm}
\end{figure}
\be
 \lagr_{int}^{NP} = i \,\frac{\pi^2 f}{g_s} \sum_\b \sum_{a,k,l}\frac{1}{m^*_\b}\,
 \bar{\psi}^{\b,k}(\eta)\, \s_{\m\n}\, t^a_{kl}\, \psi^{\b,l}(\eta)\, G_a^{\m\n}(\eta) \, ,
 \label{chrom}
\ee
 where $ f = \pi^2 n_c \r_c^{\,4}$ is the so-called packing fraction of instantons in the
QCD vacuum, $n_c$ is the instanton density, $\rho_c$ is the average instanton size,
$m^{*}_u\approx m^{*}_d\approx m^{*}$ is some appropriate mass parameter for the quark
propagator in the instanton vacuum \cc{shuryak} and
$\sigma_{\mu\nu}=[\gamma_\mu,ù\gamma_\nu]/2$. In mean field approximation for free quarks
$m^{*}$ has the meaning of an effective quark mass (see, e.g., the recent discussion in
Ref.~\cite{faccioli}).
 For off-shell quarks with virtualities $k_1$ and $k_2$ and a gluon with virtuality $q$ the instanton
induced quark-gluon vertex should be multiplied by the factor \be F(k_1,k_2,q) =
\Phi(k_1\rho_c/2)\Phi(k_2\rho_c/2)G(q\rho_c) \ , \label{form1} \ee where \ba
\Phi(z)&=&-z\frac{d}{dz}\[I_0(z)K_0(z)-I_1(z)K_1(z)\] \ , \nonumber\\
G(z)&=&\frac{4}{z^2}-2K_2(z) \label{form2} \ea are the Fourier-transformed quark zero-mode
and instanton fields respectively and $I_{\nu}(z)$, $K_{\nu}(z)$, are the well-known modified
Bessel functions. The main feature of the instanton vertex, Eq.~(\ref{chrom}), in comparison
with the perturbative one, Eq.~(\ref{lintP}), is the presence of a quark spin-flip
contribution. Moreover, the strong coupling constant $g_s$ enters the denominator of the
effective strength in Eq.~(\ref{chrom}); therefore there is not any $\alpha_s$ power
suppression  of the spin-flip contributions in our nonperturbative approach to SSA. We should
also mention that the average instanton size is much smaller than the confinement radius,
$\rho_c\approx (1/3)R_{conf}$, so that the instanton contribution to the SSA should be
significant at larger momentum transfer in comparison with the contribution coming from quark
bound-state dynamics.

Let us now estimate the contribution of the chromomagnetic interaction to the quark Sivers
function coming from Fig.~1b. Inserting Eq.~(\ref{chrom}) and the gauge links,
Eq.~(\ref{linkx}), into Eq.~(\ref{basic}), and taking the leading term we find

\ba \sivfa{\alpha}_I &=& -i\frac{M_PE_P}{k_y} \frac{\pi^2 f}{m^*} \
\intksi\  e^{-i k\cdot\xi} \int\! d^3 \vecc r \int_{\xi^{-}}^{\inf} dz^{-} \int d^4\eta \nonumber\\
&\times& \<\,\uparrow_x|\, \bar{\psi}^{\a,i}(\xi+\vecc{r})
A^{+}_{a}(z^{-},\x_{\perp}+\vecc{r})t^{a}_{ij}
\gamma^{+}\psi^{\a,j}(\vecc{r})\nonumber\\
&&\ \times  \sum_\beta \bar{\psi}^{\beta,k}(\eta)\sigma_{\mu\nu} t^{b}_{kl}
\psi^{\beta,l}(\eta)[\partial^{\mu}\!A^{\nu}_{b}(\eta)-\partial^{\nu}\!A^{\mu}_{b}(\eta)]|\uparrow_x\,\>
+ h.c. \ \ . \label{nonsimple0} \ea

As one can see, the dependence on the strong coupling constant disappears and the instanton
contribution to the Sivers function is proportional to the packing fraction of instantons in
the vacuum. We now use Eq.~(\ref{psi}) for the quark fields; by taking also the Fourier
transforms of the gluon fields, using the perturbative gluonic propagator (given consistently
in a covariant gauge, see comments after Eq.~(\ref{linkx})\,) as an approximation for the
correlator of two gauge fields, \be \<\tilde A^+_a (q) \tilde A^\n_b (p) \> = (2\pi)^4 \
\d^{(4)} (q+p) \ \(\frac{-i g^{+\n}\d_{ab}}{q^2} \) \ , \ee and using Eq.~(\ref{form1}) where
appropriate, we get

\ba \sivfa{\alpha}_I &=& -\frac{M_PE_P}{k_y} \frac{\pi^2 f}{m^*} \ \int\frac{d^4
q}{(2\pi)^4}\, G(q\rho_c)\,\frac{1}{q^2}\,\frac{1}{q^{+}+i\epsilon}\int\frac{d^3\vecc
k_1}{(2\pi)^3}\int
\frac{d^3\vecc k_3}{(2\pi)^3}\nonumber\\
&\times&2\pi\delta(q^0)\delta(k^{+}-q^{+}-k_{1}^{+})\delta^{(2)}(\vecc{k}_{\perp}-\vecc{q}_{\perp}-
\vecc{k}_{1\perp})\nonumber\\
&\times& \sum_{m_i}
\sum_{\beta}\sum_{aijkl}t^{a}_{ij}t^{a}_{kl}\<\,\uparrow_{x}|\,a^{i\,\dag}_{\alpha, m_1}
a^{j}_{\alpha, m_2}a^{k\,\dag}_{\beta, m_3}a^{l}_{\beta, m_4}|\uparrow_{x}\,\> \\
&\times& \phi^{\dag}_{m_1}(\vecc k_1)\gamma^{0}\gamma^{+}\phi_{m_2}(\vecc{k}_1+\vecc{q})
\phi^{\dag}_{m_3}(\vecc k_3)\gamma^{0}(\not\! q\, \gamma^{+}-\gamma^{+}\!\! \not\! q)
\phi_{m_4}(\vecc{k}_3-\vecc{q}) + h.c. \ \ . \nonumber \label{nonsimple1} \ea

Concerning the hermitian conjugate contribution, one can easily see that it can be obtained
by simply exchanging $q \leftrightarrow -q$ in the first term of the previous equation.
Taking also into account the relation
 \be \frac{1}{q^+ - i\e} -  \ \frac{1}{q^+ + i\e} = i (2\pi) \d(q^+)
\ , \ee performing the integrations over coordinates, and making use of the Dirac delta
functions, one can simplify the sum of the two contributions, which finally reads
 \ba
\sivfa{\alpha}_I &=& i\,\frac{M_PE_P}{k_y} \frac{\pi^2 f}{m^*}\,\sqrt{2} \ \int\!
\frac{d^2{\vecc q}_\perp G(q_\perp\,\rho_c)}{(2\pi)^2} \into  \intth \ \frac{1}{\vecc
q_\perp^2}\nonumber\\
 &\times& \d(k^+ -k_1^+ -q^+)\,
\d^{(2)}(\vecc{k}_\perp - \vecc{k}_{1\perp} - \vecc{q}_\perp)\,\sum_{m_i}\ {T}_{ \{m\} }^\a\ \nonumber\\
 &\times& \phi^{\dag}_{m_1}(\vecc k_1)\gamma^{0}\gamma^{+}\phi_{m_2}(\vecc{k}_1+\vecc{q}_\perp)\,
 \phi^{\dag}_{m_3}(\vecc k_3)\Gamma(\vecc{q}_\perp)\phi_{m_4}(\vecc{k}_3-\vecc{q}_\perp)
 \ , \label{fI}
\ea where $q_\perp=|\,\vecc{q}_\perp|$,
 \begin{eqnarray} &&{T}_{ \{m\} }^\a = \sum_{\beta}\sum_{aijkl}t^{a}_{ij}t^{a}_{kl}
 \<\,\uparrow_{x}|\,a^{i\,\dag}_{\alpha, m_1}a^{j}_{\alpha, m_2}
 a^{k\,\dag}_{\beta, m_3}a^{l}_{\beta, m_4}|\uparrow_{x}\,\> \label{tamgen}\\
 &=&\frac{1}{2}\,\sum_{\beta}\sum_{aijkl}t^{a}_{ij}t^{a}_{kl}
 \bigl(\,\<S_{+z}|\,a^{i\,\dag}_{\alpha, m_1}a^{j}_{\alpha, m_2}
 a^{k\,\dag}_{\beta, m_3}a^{l}_{\beta, m_4}|S_{-z}\>+
 \<S_{-z}|\,a^{i\,\dag}_{\alpha, m_1}a^{j}_{\alpha, m_2}
 a^{k\,\dag}_{\beta, m_3}a^{l}_{\beta, m_4}|S_{+z}\>\,\bigr)\,,\nonumber
 \label{gammaq}
\end{eqnarray}
 \be \G(\vecc{q}_\perp)=\gamma^{0}\,(\vecc{\gamma}\cdot\vecc{q}_\perp\ \gamma^{+}-\gamma^{+}
\vecc{\gamma}\cdot\vecc{q}_\perp) \ , \ee and  where $\alpha$ (not summed), $\beta$ $=u,d $
are flavour indexes and $\{m\}$ stands for the set of quark helicity indexes $(m_1, m_2, m_3,
m_4)$.

It is useful to compare this result with the perturbative contribution, Ref.~\cc{YUAN}:
 \be
 \sivfa{\alpha}_{pQCD} =  i\,2g_s^2 \,\frac{M_PE_P}{k_y}\,\sum_{m_i}\ T_{\{m\}}^\a
  \int\! \frac{d^2{\vecc q}_\perp}{(2\pi)^5}\,\frac{1}{\vecc{q}^2_\perp}\,
  F^s_{P\ m_3,m_4}(\vecc{q}_\perp^2)\,F^c_{m_1,m_2}(\vecc{q}_\perp,\vecc{k})\, ,\label{fpQCD}
 \ee
 where the current quark-gluon vertex contribution (common to both the perturbative and the instanton
 approach) is given by
\be
 F^c_{m_1,m_2}(\vecc{q}_\perp,\vecc{k})=
 \f^{\dag}_{m_1}(\vecc{k}-\vecc{q}_\perp)\gamma^0\gamma^+\f_{m_2}(\vecc{k})\, ,
\ee
 and consists of two terms, with and without helicity flip of the struck quark:
 \begin{eqnarray}
\ F_{m_1,m_2}^c (\vecc{q}_\perp,\vecc{k}) &=& \frac{1}{\sqrt{2}}\,\Bigl(\,
F^{c;\,even}_{m_1}\,\d_{m_1, m_2} + F^{c;\,odd}_{m_1}\, \d_{m_1,-m_2}\,\Bigr) \nonumber\\
&=& \frac{1}{\sqrt{2}}\ \tilde
 N^2\,\Biggl\{\,\Bigl[\,t_0(k)t_0(k')+\frac{k_z}{k}\,t_1(k)t_0(k')
 + \frac{k_z}{k'}\,t_0(k)t_1(k') \nonumber\\
 &+& \frac{1}{kk'}\,\bigl[\,\vecc{k}\cdot\vecc{k}'+i\,2m_1(\,k_x\,q_{\perp\,y}-k_y\,q_{\perp\,x})\,\bigr]
 \,t_1(k)t_1(k')\,\Bigr]\,\d_{m_1, m_2}\nonumber\\
 &+&\Bigl[\,\frac{(2m_1\,k_x-i\,k_y)}{k}\,t_1(k)t_0(k')-\frac{(2m_1\,k_x'-i\,k_y')}{k'}\,
 t_0(k)t_1(k')\nonumber\\
 &+& \frac{k_z(2m_1\,q_{\perp x}-i q_{\perp y})}{kk'}\,t_1(k)t_1(k')\,\Bigr]\,\d_{m_1,-m_2}\,\Biggr\}
 \,,\label{current}
 \end{eqnarray}
where $\tilde{N}^2=4\pi N^2 R_0^6$, $\,k=|\,\vecc{k}\,\!|$,
$\,\vecc{k}'=\vecc{k}-\vecc{q}_\perp$, $\,k'=|\,\vecc{k}'\,\!|$. Notice that both
$F^{c;\,even}_{m_1}$ and $F^{c;\,odd}_{m_1}$ have a symmetric, $F^c_{sym}$, and an
antisymmetric, $F^c_{antisym}$, component for $m_1 \leftrightarrow -m_1$.

The (perturbative) spectator quark-gluon vertex contribution reads
 \be F^s_{P\ m_3,m_4}(\vecc{q}_\perp^2) =
  \int\frac{d^3\vecc{k}_3}{(2\pi)^3}\ \f^{\dag}_{m_3}(\vecc{k}_3)\gamma^0
\gamma^+\f_{m_4}(\vecc{k}_3-\vecc{q}_\perp) = \frac{1}{\sqrt{2}}\
F^{s}_{P}(\vecc{q}_\perp^2)\,\delta_{m_3,m_4}\, \label{pert-spec}\ee where the spectator
function $F^s_P(\vecc{q}_\perp^2)$ is given by (see also Ref.~\cite{YUAN})

\begin{eqnarray}
F^{s}_{P}(\vecc{q}_\perp^2) &=& \frac{\tilde{N}^2}{(2\pi)^2}\,\int \k^2\, d\k\,
d(\cos\,\theta)\,\biggl\{\,t_0(\k)t_0(\k')+\frac{\k\,\cos\,\theta-q_\perp}{\k'}\,t_0(\k)t_1(\k')
\nonumber\\
&+& \cos\,\theta\,t_1(\k)t_0(\k')+\frac{\k-q_\perp
\cos\,\theta}{\k'}\,t_1(\k)t_1(\k')\,\biggr\}\,, \label{fsp}
\end{eqnarray}
and we have defined $\vecc{\k}=\vecc{k}_3$, $\,\k=|\,\vecc{k}_3\,\!|$,
$\,\vecc{\k}'=\vecc{k}_3-\vecc{q}_\perp$, $\,\k'=|\,\vecc{\k}'\,\!|=
 \bigl[\,\k^2+q_\perp^2-2\k\,q_\perp \cos\,\theta\,\bigr]^{1/2}$\,.

It is worth to emphasize at this point that, according to our definition of the matrix
elements $\<\uparrow_{x}|\dots|\uparrow_{x}\>$ (see comments after Eq.~(\ref{basic}) and
Eq.~(\ref{tamgen})\,) the only non-vanishing $T^\alpha_{m_1,m_2,m_3,m_4}$ are those of the
form $T^\alpha_{m_1,m_1,m_3,-m_3}$ or $T^\alpha_{m_1,-m_1,m_3,m_3}$. Therefore, one-gluon
exchange contributions, associated either with the perturbative or the nonperturbative
spectator vertex, can be non-vanishing only when we have just (and at least) one
helicity-flip term, either in the struck quark vertex, or in the spectator one, but not in
both of them. In the perturbative case, the {\it struck-quark helicity-flip} situation takes
place, Ref.~\cc{YUAN}. As we will see below, the specific instanton-induced spin-spin
correlation allows both the {\it struck-} and {\it spectator-quark helicity flip} mechanisms
to contribute.

In the nonperturbative case, it is convenient to define an instanton analogue of the function
$F^{s}_{P}(\vecc{q}_\perp^2)$:
 \begin{eqnarray}
 F^s_{I\ m_3,m_4} (\vecc{q}_\perp) &=& \int \frac{d^3\vecc{k}_3}{(2\pi)^3}\
 \phi^{\dag}_{m_3}(\vecc k_3)\Gamma(\vecc{q}_\perp)\phi_{m_4}(\vecc{k}_3-\vecc{q}_\perp)
 \nonumber\\  &=& \frac{1}{\sqrt{2}}\,
 \Bigl(\,F^{s;\,even}_{I\ m_3}(\vecc{q}_\perp)\,\delta_{m_3,m_4}
  + F^{s;\,odd}_{I\ m_3}(\vecc{q}_\perp)\,\delta_{m_3,-m_4}\,\Bigr)\, ,
  \label{fsi}
\end{eqnarray}

so that the instanton contribution to the Sivers function reads
 \be \sivfa{\a}_I = i\frac{M_P
E_P}{k_y} \frac{\pi^2 f}{m^*}\,2 \int \frac{d^2\vecc{q}_\perp}{(2\pi)^5} \frac{1}{
\vecc{q}_\perp^2}\,G(q_\perp\,\rho_c)\,\sum_{m_i} {T}_{\{m\}}^\a\,
 F_{m_1,m_2}^c (\vecc{q}_\perp,\vecc{k}) \ F_{I \ m_3,m_4}^s (\vecc{q}_\perp) \label{split} \,. \ee

 Working in the standard representation of the Dirac matrices \be \vecc \g = \({\ \  0 \ \ \ \ \vecc
\s}\atop{-\vecc \s \ \ 0}\) \ , \  \g^0\vecc \g \g^+ = \frac{1}{\sqrt 2} \({-\vecc \s \s_3 \
 -\!\vecc \s}\atop{\ \ \ \vecc \s \ \ \ \ \vecc \s \s_3}\) \ , \ \g^0\g^+\vecc \g = \frac{1}{\sqrt
2} \({-\s_3 \vecc \s \ \ \vecc \s}\atop{- \vecc \s \ \ \s_3 \vecc \s} \) \ , \ee
 one finds
\begin{eqnarray}
F^s_{I\ m_3,m_4}(\vecc{q}_\perp)\!\!\! &=& \!\!\!\!\sqrt{2}\, \tilde N^2 \int
\frac{d^3\vecc{\k}}{(2\pi)^3}\, \c^\dag_{m_3}\, \{\,i [\vecc{q}_\perp \times \vecc \s]_z \
t_0(\k)t_0(\k') - (\vecc \s
\!\cdot\!\vecc{q}_\perp) (\vecc \s\!\cdot\!{\vecc{\k}'})\ \frac{1}{\k'}\,t_0(\k)t_1(\k')\nonumber\\
\!\!\!&+& \!\!\!\!(\vecc \s\!\cdot\!{\vecc{\k}})(\vecc \s \!\cdot\!\vecc{q}_\perp)\
\frac{1}{\k}\,t_1(\k)t_0(\k') - i (\vecc \s \!\cdot\!{\vecc{\k}}) [\vecc{q}_\perp \times
\vecc \s]_z (\vecc \s\!\cdot\!{\vecc{\k}'})\ \frac{1}{\k\k'}\,t_1(\k)t_1(\k')
\,\}\,\c_{m_4}\, . \nonumber \\ \label{fsi-ch}
\end{eqnarray}

After some algebra and taking into account symmetry properties of the integrand, one gets
\begin{eqnarray}
 F^{s;\,even}_{I\ m_3} (\vecc{q}_\perp) \!\!\!&=&\!\!\! \frac{2\tilde{N}^2}{(2\pi)^2}\,\int\! \k^2 d\k\
 d(\cos\,\theta)\,q_\perp\(\,\cos\,\theta \,t_0(\k') t_1(\k) - \frac{\k\,\cos\,\theta - q_\perp}{\k'}
 \,t_0(\k)t_1(\k') \)\ \ \ \  \label{eveninst} \\
F^{s;\,odd}_{I\ m_3} (\vecc{q}_\perp) \!\!\!&=&\!\!\! \frac{2\tilde{N}^2}{(2\pi)^2}\
(2m_3\,q_{\perp x}-iq_{\perp y})\, \int\! \k^2 d\k\ d(\cos\,\theta) \ \nonumber\\
&\times& \(t_0(\k) t_0(\k') + \frac{\k - q_\perp\,\cos\,\theta}{\k'} \ t_1(\k)t_1(\k') \)\, .
\label{oddinst}
\end{eqnarray}

Notice that $F^{s;\,even}_{I\ m_3}$ is in fact independent of $m_3$, while $F^{s;\,odd}_{I\
m_3}$ has both a symmetric and an antisymmetric component, $F^{s;\,odd}_{I\ sym}$ and
$F^{s;\,odd}_{I\ antisym}$, for $m_3 \leftrightarrow -m_3$.

This time we can have two additive contributions to the Sivers function:
$F^{c;\,odd}_{s;\,even}$ and $F^{c;\,even}_{s;\,odd}$, corresponding respectively to
helicity-flip of the struck quark and no helicity flip for the spectator quark, and {\it
viceversa},
 \be
 \sivfa{\alpha}_{I} \propto \sum_{m_1,m_3}\,\Bigl(\,F^{c;\,even}_{m_1}\,F^{s;\,odd}_{m_3}\,
 T^{\alpha}_{m_1,m_1,m_3,-m_3}+F^{c;\,odd}_{m_1}\,F^{s;\,even}_{m_3}\,
 T^{\alpha}_{m_1,-m_1,m_3,m_3}\,\Bigr)\,.\label{odd-even}
 \ee
 Therefore both the spectator and the current quark spin-flip terms
contribute to the instanton-induced Sivers function. This is in contrast to the purely
perturbative result, Eq.s~(\ref{current}) and (\ref{pert-spec}), where only the current quark
spin-flip gives a non-vanishing contribution to the Sivers function. Moreover, as we will see
in a moment, it turns out that the instanton contribution has a completely different flavour
dependence, due to the structure of the matrix element in  Eq.~(\ref{split}). This could open
a new way of understanding the unexpected flavor dependence of the Sivers function, as
extracted from recent  HERMES data \ct{HERMES2} (see below).

An explicit calculation of the non-vanishing proton matrix elements $T^{\alpha}_{\{m\}}$ in
Eq.~(\ref{tamgen}) shows that they have the same value for any allowed combination of $m_1$,
$m_3$, so that, e.g.,
 \begin{eqnarray}
 F^{c;\,even}_{s;\,odd} &=& \sum_{m_1,m_3} F^{c;\,even}_{m_1}\,F^{s;\,odd}_{I\ m_3}\,
 T^{\alpha}_{m_1,m_1,m_3,-m_3}
 = T^{\alpha}_{\ \uparrow\,\uparrow\,\uparrow\,\downarrow}
 \sum_{m_1,m_3} F^{c;\,even}_{m_1}\,F^{s;\,odd}_{I\ m_3}
 \nonumber\\
 &=& \sum_{m_1,m_3} T^{\alpha}_{m_1,m_1,m_3,-m_3}\,\Bigl(\,\frac{1}{2}\,
 \sum_{m_1}\,F^{c;\,even}_{m_1}\,\Bigr)\,\Bigl(\,\frac{1}{2}\,
 \sum_{m_3}\,F^{s;\,odd}_{m_3}\,\Bigr) \nonumber\\
 &=& C^\a_I|^{\,even}_{\,odd}\, F^{c;\,even}_{sym} \, F^{s;\,odd}_{I\ sym}\, ,
 \label{feo}
 \end{eqnarray}
where $F^{c;\,even}_{sym}$ and $F^{s;\,odd}_{I\ sym}$ can be obtained from
Eq.s~(\ref{current}) and (\ref{oddinst}) respectively. Analogously,
 \be
 F^{c;\,odd}_{s;\,even} = C^\a_I|^{\,odd}_{\,even}\, F^{c;\,odd}_{sym} \, F^{s;\,even}_{I}\, ,
 \label{foe}
 \ee
where again $F^{c;\,odd}_{sym}$ can be obtained from Eq.~(\ref{current}).

Notice that $C^\a_I|^{\,odd}_{\,even}$ corresponds to the perturbative case, see Eq.s
(\ref{fpQCD}), (\ref{current}), (\ref{pert-spec}) and Ref.~\cite{YUAN}. Straightforward
calculation gives \be C^u_I|^{\,even}_{\,odd} = - \frac{4}{9} \ \ , \ \
 C^d_I|^{\,even}_{\,odd} = - \frac{8}{9} \label{even-odd} \ee and
  \be C^u_I|^{\,odd}_{\,even} = - \frac{16}{9} \ \ , \ \
   C^d_I |^{\,odd}_{\,even} = + \frac{4}{9} \ . \label{odd-even1} \ee

Therefore, in the instanton approach the total expression for the Sivers function has two
contributions of a very different origin, which sum up together:
\begin{eqnarray}
 &&\sivfa{\a}_I = M_P E_P\,\frac{f}{2m^*}\,\tilde{N}^4\,\frac{1}{(2\pi)^5} \int
 d^2\vecc{q}_\perp\, \frac{1}{
 \vecc{q}_\perp^2}\,G(q_\perp\,\rho_c)\qquad\qquad\qquad\qquad\qquad\nonumber\\
 &\times\ &\!\!\!\!\!\! \Biggl\{\ C^\a_I|^{\,even}_{\,odd}\,\biggl[\,t_0(k)t_0(k')+\frac{k_z}{k}\,
 t_1(k)t_0(k')+\frac{k_z}{k'}\,t_0(k)t_1(k')+
 \frac{k^2-k_\perp\,q_\perp\,\cos\,\phi}{kk'}\,t_1(k)t_1(k')\,\biggr]\nonumber\\
 &&\!\!\!\! \times\ \frac{q_\perp\,\cos\,\phi}{k_\perp}\int\! \k^2 d\k\ d(\cos\,\theta)
 \biggl[\,t_0(\k) t_0(\k') + \frac{\k - q_\perp\,\cos\,\theta}{\k'} \ t_1(\k)t_1(\k')\,\biggr]
 \label{final}\\
 &+&C^\a_I|^{\,odd}_{\,even}\,\biggl[\,\frac{k_\perp}{k}\,t_1(k)t_0(k')-
 \frac{k_\perp-q_\perp\,\cos\,\phi}{k'}\,t_0(k)t_1(k')+
 \frac{k_z\,q_\perp\,\cos\,\phi}{k k'}\,t_1(k)t_1(k')\,\biggr]\nonumber\\
 &&\!\!\!\! \times\ \frac{q_\perp}{k_\perp} \int\! \k^2 d\k\,d(\cos\,\theta)\,
 \biggl[\,\cos\,\theta \,t_0(\k') t_1(\k) -
 \frac{\k\,\cos\,\theta -q_\perp}{\k'}\,t_0(\k)t_1(\k')\,\biggr]\ \Biggr\}
 \, ,\nonumber
\end{eqnarray}
where $\phi$ is the azimuthal angle of $\vecc{q}_\perp$, and
 \be
  k_z=\frac{\omega(4x-1)}{R_0}\,,\quad
  k=\sqrt{k_z^2+k_\perp^2}\,,\quad
  k'=\sqrt{k_z^2+(\vecc{k}_\perp-\vecc{q}_\perp)^2}\, .
  \label{kz}
 \ee

Notice that to go from Eq.s~(\ref{current}), (\ref{eveninst}), (\ref{oddinst}), (\ref{split})
to Eq.~(\ref{final}) we have used the identity

\be
 \int d^2\vecc{q}_\perp\,q_{\perp i}\,F(\vecc{k}^2,\vecc{q}_\perp^2,
 \vecc{k}_\perp\!\cdot\!\vecc{q}_\perp) \equiv \frac{k_i}{k_\perp^2} \int d^2\vecc{q}_\perp\,
 \vecc{k}_\perp\!\cdot\!\vecc{q}_\perp\,F(\vecc{k}^2,\vecc{q}_\perp^2,
 \vecc{k}_\perp\!\cdot\!\vecc{q}_\perp)\, ,
  \label{fequiv}
\ee where $F$ is any function with the functional dependence shown and $i=x,y$. In this way,
we can effectively take in Eq.s~(\ref{current}), (\ref{eveninst}), (\ref{oddinst}), $q_{\perp
i} \rightarrow k_i\,(\vecc{k}_\perp\!\cdot\!\vecc{q}_\perp)/k_\perp^2$, and the numerator of
the Sivers function, Eq.~(\ref{split}), becomes proportional to $k_y$, which simplifies, as
it must be, with the overall factor $1/k_y$ originally present in our expression.

\section{Numerical results}

In order to estimate the Sivers function for $u$ and $d$ valence quarks within the instanton
approach proposed in this paper, we do not make any attempt to fix the parameters of the
instanton model by fitting available results on single spin asymmetries. Rather, we consider
widely accepted values for these parameters, coming from a comprehensive application of the
instanton model to nonperturbative QCD (see Ref.~\cite{shuryak} for more detail): $\r_c
\approx 0.3$ fm, $m^*\approx 170$ MeV and $n_c \approx 0.5$ fm$^{-4}$.
 For the perturbative contribution we will take the following value of the strong coupling constant,
$\alpha_s\approx 0.3$. This value  is usually adopted for estimation of the one-gluon
exchange contribution to SSA for HERMES kinematics (see e.g. Ref.~\cite{bacchetta}). This has
to been taken into account when comparing our results with those of Yuan \cite{YUAN}.
\begin{figure}
%\centering
\begin{minipage}[c]{7.5cm}
\vspace*{-0.92cm}
%\centering
\hspace*{0.0cm} \epsfig{file=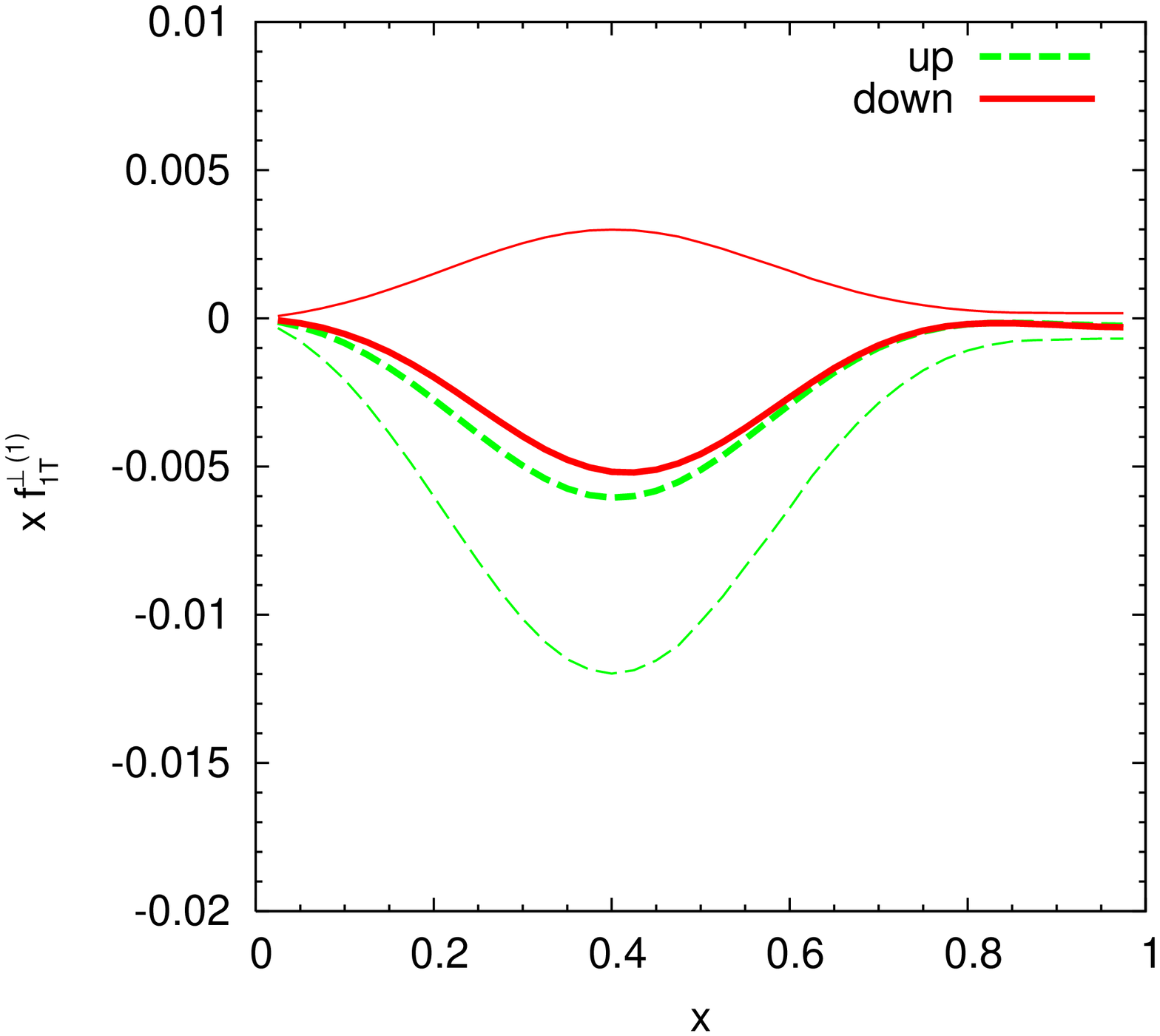,width=6.5cm,angle=-90}
%\vskip -8cm
\caption{The instanton (thick lines) and the one-gluon (thin lines) contributions to the
first moment of the $d$-quark (solid) and $u$-quark (dashed) Sivers function, vs. $x$.}
%\vspace*{-5cm}
\end{minipage}
\hspace*{0.5cm}
\begin{minipage}[c]{7.5cm}
%\centering
\vspace*{-1.0cm}
 \hspace*{0.0cm} \epsfig{file=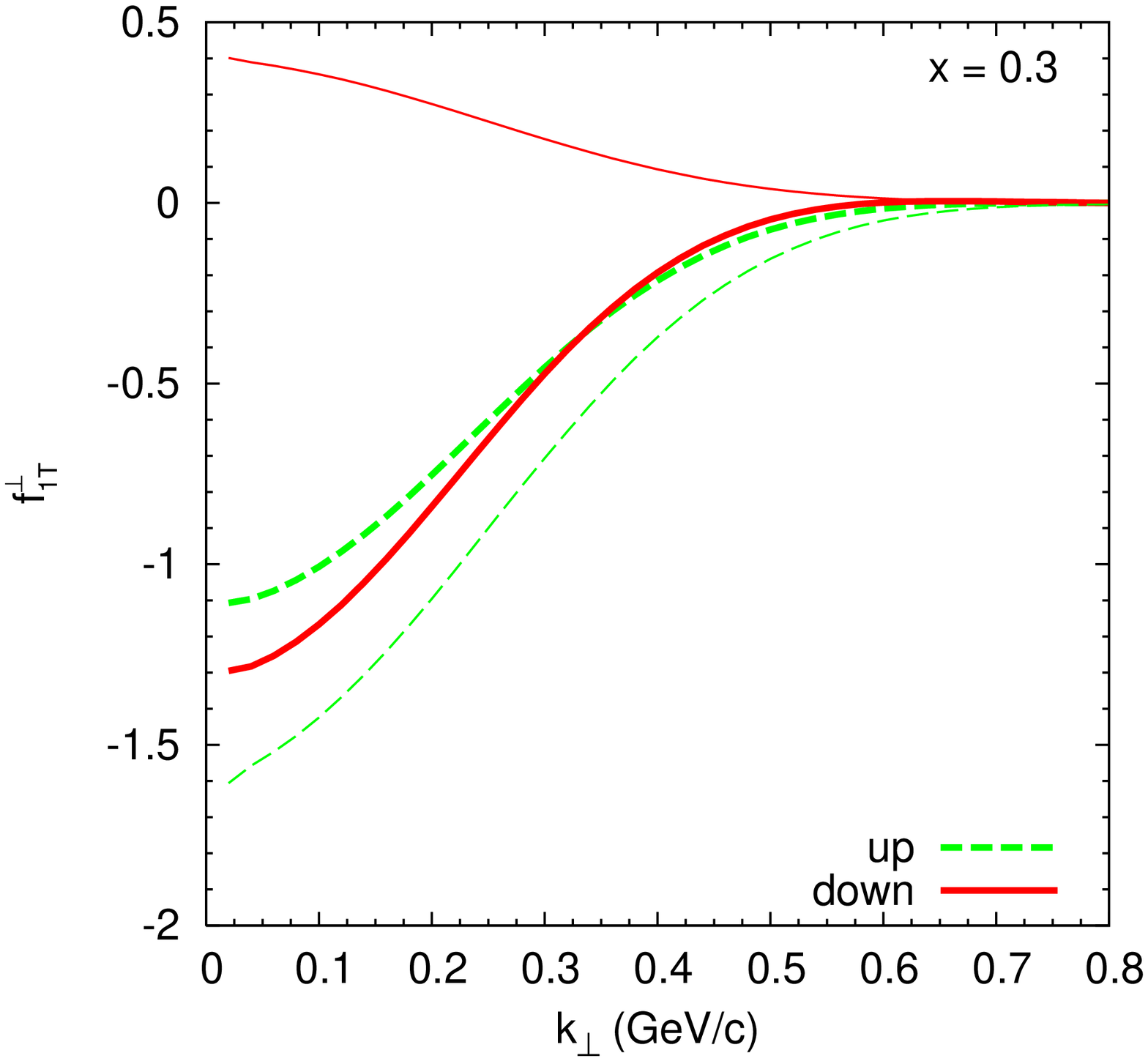,width=6.5cm,angle=-90}
%\vspace*{0.0cm}
 \caption{The instanton and one-gluon contributions to the Sivers function at
$x$=0.3 vs. $k_\perp$(GeV$/c$). Notations are the same as in Fig.~2.}
\end{minipage}
\end{figure}
\begin{figure}

\centering
%\centering
\hspace*{0.3cm}
  \epsfig{file=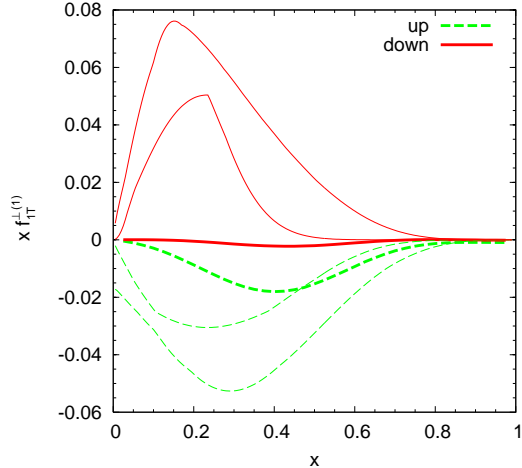,width=6.5cm,angle=-90}
%\vskip -8cm
\caption{The total contribution to the first moment of $d$ (solid line) and $u$ (dashed line)
quark  Sivers distributions as a function of $x$. The 1-$\sigma$ band for the
parameterizations extracted by fitting the HERMES data (see first of Ref.s~\cite{sidis}) is
also shown.}
%\vspace*{-5cm}

\end{figure}

Using these parameters, in Fig.~2 we present  the instanton and the perturbative, one-gluon
exchange, contributions to the first moment of the Sivers function,
 \be
 f_{1\,T}^{\perp(1)\,\alpha}(x) = \int d^2\vecc{k}_{\perp}\,\(\frac{k_{\perp}^2}{2M_P^2}\)\,
 f_{1\,T}^{\perp\,\alpha}(x,k_{\perp})\,, \label{mom}
 \ee
for $u$ and $d$ quarks, as a function of $x$. In Fig.~3, the Sivers function is shown as a
function  of $k_\perp$, at $x=0.3$. The main features of these results are that the instanton
contribution is sizable and, in contrast with the purely perturbative case, it has the same
(negative) sign and approximately the same magnitude for $u$ and $d$ quarks.

As a consequence, when summing the two independent perturbative and instanton terms to get
the total contribution to the Sivers function, for the $d$-quark the two terms almost cancel
each other, leading to a small, negative Sivers function. On the contrary, in the case of the
$u$-quark Sivers function, the instanton contribution sums up with the perturbative one. This
can be seen in Fig.~4, where we also compare the total contribution to the first moment of
the Sivers function with recent parameterizations obtained by fitting the results of the
HERMES Collaboration for the single spin asymmetry in semi-inclusive DIS pion production
\cite{HERMES2}, see first of Ref.s~\cite{sidis}. {}From these results, one can observe that
the present estimate of the instanton contribution to the valence quark Sivers function
compares reasonably well, for $u$ quarks, with parameterizations leading to a good
description of the HERMES data in the valence-quark region, $x \geq~0.3$, where one can
believe in the MIT-bag based calculations. On the other hand, for $d$ quarks our results are
smaller than one could expect from data fitting and differ from those of other model
calculations and available parameterizations.

\section{Conclusions}

In this paper, the instanton contribution to the valence quark Sivers function has been
estimated, adopting the MIT bag model for the nucleon wave function. Our results show that
this contribution can be sizable, since it is not suppressed by powers of the strong coupling
constant. The specific flavour dependence of the instanton contribution, as compared to that
of the perturbative one, leads to a large, negative total contribution to the $u$-quark
Sivers function. On the other hand, it suppresses the total contribution in the $d$-quark
case. However, several points should be kept in mind here:\\
a) We have used widely accepted values for the basic parameters of the instanton model. These
parameters are known with relatively large uncertainties which reflect into the results of
our calculations. As an example, using the Diakonov and Petrov instanton liquid model
estimate for the quark anomalous chromomagnetic moment (see e.g. Ref.~\cite{diakonov}) would
lead to an almost twice larger instanton contribution.\\
 b) The value adopted for the strong coupling constant plays a role in the interplay between
 the purely perturbative and the instanton terms, since the first contribution is
 proportional to $\alpha_s$ while the second one depends on it in a more involved and
 indirect way. Changing the effective value of $\alpha_s$, which is not well known a priori, would
 therefore modify in particular the $d$ quark case.\\
 c) Due to the effective instanton size, the instanton contribution could in principle lead to a
 different $k_\perp$ dependence of the Sivers function as compared to the purely perturbative case.
 In our results, however, the two contributions show a very similar behaviour in this
 respect. This is because in this first simplified approach the effective $k_\perp$
 dependence of the Sivers function is basically determined by the MIT-bag model wave functions
 rather than by specific instanton properties.\\
 d) There could be in principle additional instanton contributions to the Sivers function, coming
 from more complex Feynman diagrams, involving e.g. the spectator diquark as a whole. Moreover, these
 contributions might be flavour-dependent, leading to different changes in the $u$ and $d$
 quark Sivers function.

 A detailed treatment of all these points is beyond the scope of this paper, which was
 mainly intended to show that instanton effects may be relevant for the Sivers distribution.
A deepest study of instanton contributions to the Sivers function, including additional
diagrams and based on more elaborated models for the nucleon wave function is in progress
\cite{prog}.

\section*{Acknowledgments}
We are very grateful to Feng Yuan for enlightening discussions. We also thank D.I. Diakonov,
A.E. Dorokhov, K. Goeke, W.-D. Nowak, P.V. Pobylitsa, M.V. Polyakov, A.V. Prokudin, A.E.
Radzhabov, and N. Stefanis for fruitful critical comments. I.O.C. is grateful to Ulf-G.
Meissner and B. Metsch for hospitality at Bonn University in the final stage of this work.
N.I.K. is grateful to the School of Physics, SNU, and especially to Prof. Dong-Pil Min for
their warm hospitality. This work is partially supported by: the Heisenberg-Landau program
and the Grant of Russian Foundation for Basic Research, No. RFBR-04-02-16445 (I.O.C. and
N.I.K.); the Russian President's Grant No. 1450-2003-2, and the Alexander von
Humboldt-Stiftung and DFG (I.O.C.); the Brain Pool program of Korea Research Foundation
through KOFST, Grant No. 042T-1-1 (N.I.K.); the European Community-Research Infrastructure
Activity under the FP6 ``Structuring the European Research Area'' programme, HadronPhysics,
Contract No. RII3-CT-2004-506078 (U.D. and F.M.).

\end{document}